\documentclass[prl,superscriptaddress,twocolumn,amsmath,amssymb,nofootinbib,longbibliography]{revtex4-2}

\usepackage{graphicx}
\usepackage{amsmath}
\usepackage{physics}
\usepackage{color}
\usepackage{xcolor}
\usepackage{slashed}
\usepackage{tensor}
\usepackage[utf8]{inputenc}
\usepackage[normalem]{ulem}

\newcommand{\ba}{\begin{eqnarray}}
\newcommand{\ea}{\end{eqnarray}}

\newcommand{\be}{\begin{equation}}             
\newcommand{\ee}{\end{equation}}               
\newcommand{\notprop}{\propto\kern-1\@ptsize pt \diagup}

\begin{document}

\title{Multicritical Phase Transitions in Lovelock AdS Black Holes}

\author{Jerry Wu}
\email{yq4wu@uwaterloo.ca} 
	\affiliation{Department of Physics and Astronomy, University of Waterloo,
		Waterloo, Ontario, Canada, N2L 3G1}
	
\author{Robert B. Mann}
	\email{rbmann@uwaterloo.ca}
		\affiliation{Department of Physics and Astronomy, University of Waterloo,
		Waterloo, Ontario, Canada, N2L 3G1}	
	\affiliation{Perimeter Institute, 31 Caroline Street North, Waterloo, ON, N2L 2Y5, Canada}

\date{\today}
	
	\begin{abstract}
		We demonstrate that black holes in order $N\ge 4$ Lovelock gravity 
can exhibit multicritical phase behaviour.
  We show an explicit example of a quadruple point in $d=10$ fourth-order Lovelock gravity and a quintuple point in $d=14$ sixth-order Lovelock gravity. We also demonstrate that multi-criticality can be realized for  uncharged, non-rotating black holes by highlighting a new type of multi-critical point between black holes and thermal radiation. We discuss the methodology used and make comparisons to other black hole multi-critical points in terms of the Gibbs phase rule. 
	\end{abstract}

\maketitle

\section{Introduction}

Despite the elegance and empirical success of  Einstein’s General Theory of Relativity  \cite{Will:2018bme}, it has yet to be reconciled with quantum theory.   A key ingredient in formulating a 
 full description of quantum gravity will be provided by black holes. Classically a black hole is a perfect absorber of matter and energy, but  once quantum physics is taken into account  its behaviour drastically changes \cite{Hawking:1975vcx}: it is predicted to radiate 
  like a thermal blackbody at a  temperature proportional to its surface gravity and with an entropy proportional to its horizon area. 
  
 In conjunction with this is another general expectation that  higher curvature terms will appear in addition to the Einstein-Hilbert action 
 due to quantum gravitational effects  \cite{Birrell:1982ix}.   The most physically sensible such generalization is
 regarded to be Lovelock gravity \cite{Lovelock:1971yv,Lovelock:1972vz} as its field equations are second order in all metric components, though higher dimensions are required to render them non-trivial.  They are of particular interest in black hole thermodynamics, since they provide insight as to how the  radiative behaviour of black holes might be modified by quantum gravitational effects \cite{Frassino:2014pha}.

An early discovery in  black hole thermodynamics was that they can undergo phase transitions: a Schwarzschild anti de Sitter (AdS) black hole
can undergo a first order phase transition to thermal AdS at sufficiently low temperature \cite{Hawking:1982dh}, known as the Hawking-Page (HP)  transition.  For the past decade it has become clear that the negative cosmological constant $\Lambda$ in AdS actually plays a much more active role  as
thermodynamic pressure  \cite{Kastor:2009wy}, from which a very broad range of chemical thermodynamic phenomena ensue \cite{Kubiznak:2016qmn}. 
Thermodynamics of Lovelock black holes have been extensively investigated over the past decade, with most studies  examining second-order (Gauss-Bonnet) and third-order Lovelock theories, where triple points \cite{Frassino:2014pha,Wei:2014hba,Hull:2021bry,Wei:2021krr,Hull:2022xew}, reentrant phase transitions \cite{Frassino:2014pha}, and isolated critical points \cite{Frassino:2014pha,Dolan:2014vba} have been observed.  Some of these phenomena have also been observed in other theories incorporating higher curvature effects \cite{Hendi:2018xuy, Dehghani:2020blz,Mir:2019ecg,Mir:2019rik,Bueno:2022res}.

Very recently multi-critical black hole phase transitions were discovered, in which several distinct phases of black holes merge at a single pressure and temperature, generalizing the black hole triple point first observed for doubly-rotating black holes \cite{Altamirano:2013uqa}.  These were first observed in four-dimensional Einstein gravity coupled to power-Maxwell theory \cite{Tavakoli:2022kmo}, and shortly afterward shown to exist for any Kerr-AdS multiply rotating black hole provided there are a sufficient number of distinct, appropriately chosen angular momenta \cite{Wu:2022bdk}.  Explicit examples of quadruple points and quintuple points have been given in both cases.
 
The number of black hole phase transitions is related to the number of thermodynamic conjugate pairs, which in power-Maxwell theory are given by the number of distinct coupling parameters and their thermodynamic conjugates  \cite{Tavakoli:2022kmo}, and in the Kerr-AdS case by the number of distinct angular momenta and their conjugate angular velocities \cite{Wu:2022bdk}.   It has been known for quite some time that Lovelock coupling constants can also be regarded as thermodynamic parameters \cite{Kastor:2010gq,Kastor:2011qp,Frassino:2014pha}, and so it is reasonable to expect that multi-critical phase transitions can also take place in Lovelock gravity.

Here we demonstrate that this expectation is indeed realized, and that vacuum (charged) black hole solutions in Lovelock gravity can indeed have multi-critical points provided the number of Lovelock couplings is sufficiently large. We present multi-critical points, where more than three distinct black hole phases coexist, in the context of Lovelock gravity and find that they behave similarly to previously discovered multi-critical points. The $P$-$T$ phase diagram shows $n-1$ distinct  first order phase transitions that merge at an $n$-tuple multi-critical point at a pressure $P_{mc}$. As the pressure increases, all phase transitions terminate at their respective critical points. Only one stable first-order phase transition exists for $P<P_{mc}$. 
If the black hole is uncharged,  we find that   multiple phase transitions arising from the extended phase space can overlap with an ``incomplete'' swallowtail in the $G$-$T$ plot, indicative of an HP transition. This results in a special type of multi-critical point between thermal radiation and $n$ black hole phases of different sizes. An HP triple point was first observed in exotic Lovelock black holes \cite{Hull:2021bry}, whose horizon curvature is not constant. The HP $n$-tuple point we observe occurs for spherically symmetric black holes with constant curvature horizons. 
 As the pressure is increased, the phase transitions between black hole phases vanish at their respective critical points, while the HP transition is seen to extend up to the maximal pressure allowed by the theory. Below the multi-critical pressure, the black hole phase transitions become unstable and the HP transition is the only stable phase transition.
We explicitly show examples of a quadruple point for a charged Lovelock black hole, as well as triple and quadruple points for uncharged Lovelock black holes. 

\section{Thermodynamics of Lovelock Gravity}

The Lagrangian of an $N$th-order Lovelock theory of gravity in $d$ spacetime dimensions is \cite{Lovelock:1971yv}
\be
\mathcal{L}=\frac{1}{16 \pi G_N}\sum_{k=0}^N \hat{\alpha}_k \mathcal{L}^{(k)} \label{lagrangian}
\ee
where $\hat{\alpha}_k$ are the $k$th order Lovelock coupling constants, and 
\be
\hat{\alpha}_0 = - 2 \Lambda, \qquad P=-\frac{\Lambda}{8 \pi G_N} = \frac{\hat{\alpha}_0}{16 \pi G_N}
\ee
where $G_N$ represents Newton's constant and $P$ is the thermodynamic pressure. The quantities
 $\mathcal{L}^{(k)}$ are Euler densities given by the contraction of $k$ powers of the Riemann tensor
\be
\mathcal{L}^{(k)}= \frac{1}{2^k} \delta_{c_1 d_1 \dots c_k d_k}^{a_1 b_1 \dots c_k b_k} R_{a_1 b_1}^{\quad c_1 d_1} \dots R_{a_k b_k}^{\quad c_k d_k}
\ee
where the generalized Kronecker-delta symbol $\delta$ is fully antisymmetric in both sets of indices. We require $d\ge 2N+1$ as $\mathcal{L}^{(N)}$  makes no contribution to the equations of motion for $d\le 2N$. The term $\mathcal{L}^{(0)}$ corresponds to the cosmological constant term, whereas $\mathcal{L}^{(1)}$ and $\mathcal{L}^{(2)}$ respectively correspond to the Einstein-Hilbert term and Gauss-Bonnet term  in the Lagrangian. As such, setting $\hat{\alpha}_k =0 $ for $k>1$ recovers General Relativity.

The action of a charged black hole follows from \eqref{lagrangian}:
\be
I=\int d^d x \frac{\sqrt{-g}}{16 \pi G_N} \left(\sum_{k=0}^N \hat{\alpha}_k \mathcal{L}^{(k)} - 4 \pi G_N F_{ab} F^{ab}\right),
\ee
with $F=dA$ being the electromagnetic tensor. The corresponding field equations are 
\be
\sum_{k=0}^N \hat{\alpha}_k \mathcal{G}_{ab}^{(k)} = 8\pi G_N \left(F_{ac} F_b^{\ c} - \frac{1}{4} g_{ab} F_{cd} F^{cd} \right) \label{fieldeq}
\ee
where  the Einstein-like tensors $\mathcal{G}^{(k)}$ are given by
\be
\mathcal{G}^{(k) a}_{\ b} = -\frac{1}{2^{k+1}} \delta^{a c_1 d_1 \dots c_k d_k}_{b e_1 f_1 \dots e_k f_k} R_{c_1 d_1} ^{\quad e_1 f_1} \dots R_{c_k d_k} ^{\quad e_k f_k},
\ee
and each independently satisfies a conservation law $\nabla _a \mathcal{G}^{(k) a}_{\ b}=0$.

Employing the  metric ansatz 
\begin{align}
    ds^2 &= -f(r) dt^2 + f(r)^{-1} dr^2 + r^2 d\Omega^2 _{(\kappa)d-2}, \nonumber \\
    F &= \frac{Q}{r^{d-2}} dt \wedge dr,
\end{align}
 for static charged spherically symmetric black holes, 
the equations of motion \eqref{fieldeq} reduce to a polynomial equation  of degree $N$   in $f$ \cite{Boulware:1985wk,Wheeler:1985nh,Wheeler:1985qd,Cai:2003kt,Camanho:2011rj,Takahashi:2011du,Castro:2013pqa}
\begin{align}\label{lovepoly}
\sum_{k=0}^{N} \alpha_k & \left(\frac{\kappa -f }{r^2} \right)^k \nonumber \\ &= \frac{16 \pi G_N M}{(d-2)\Sigma^{(\kappa)}_{d-2}r^{d-1}} - \frac{8 \pi G_N Q^2}{(d-2)(d-3)r^{2d-4}}
\end{align}
where 
\begin{align}
\alpha_0 &= \frac{\hat{\alpha}_0}{(d-1)(d-2)}, \quad \alpha_1=\hat{\alpha}_1,\nonumber \\ 
\alpha_k &= \hat{\alpha}_k \prod_{n=3}^{2k} (d-n) \text{ for } k\ge2,
\end{align}
are the  rescaled Lovelock coupling constants.   
We shall consider only   spherical horizon geometries,  for which  $\kappa = +1$, and so 
\be
\Sigma^{(+1)}_{d-2} = \frac{2\pi ^{(d-1)/2}}{\Gamma(\frac{d-1}{2})}.
\ee
is the area of a $(d-2)$-dimensional unit sphere.

The characteristics of a black hole solution, including its mass $M$, temperature $T$, entropy $S$, and charge $Q$, are related by the extended first law of black hole thermodynamics \cite{Jacobson:1994qe} and the generalized Smarr relation \cite{Kastor:2010gq}
\be
\delta M = T \delta S - \frac{1}{16 \pi G_N} \sum_k \hat{\Psi}^{(k)} \delta \hat{\alpha}_k + \Phi \delta Q
\ee
\be
 M = \frac{d-2}{d-3} T S + \sum_k 2\frac{k-1}{d-3} \frac{\hat{\Psi}^{(k)} \hat{\alpha}_k}{16 \pi G_N} + \Phi Q.
\ee
where 
$\Phi$ denotes the electromagnetic gauge potential, and the potentials $\hat{\Psi}^{(k)}$ are thermodynamically conjugate to the Lovelock coupling constants $\hat{\alpha}_k$ \cite{Kastor:2010gq}.  The entropy 
\be
S= \frac{1}{4 G_N} \sum_k \hat{\alpha}_k \mathcal{A}^{(k)}, \qquad \mathcal{A}^{(k)} = k \int_\mathcal{H} \sqrt{\sigma} \mathcal{L}^{(k-1)}.
\ee
is no longer directly proportional to the horizon area as in Einstein gravity, but rather depends on higher order curvature terms. 
Here, $\sigma_{ab}$ is the induced metric on the black hole horizon $\mathcal{H}$ on which the $\mathcal{L}^{(k)}$ terms are evaluated.

The thermodynamic pressure is proportional to the zeroth-order rescaled coupling parameter
\be
P=\frac{(d-1)(d-2) \alpha_0}{16 \pi G_N}.
\ee
which in general has a maximal value if we require that a solution to  \eqref{lovepoly} yields a solution to general relativity in the limit
that all $\alpha_k \to 0$ for $k\ge 2$, or in other words, that an Einstein branch exists. The Einstein branch is given by the smallest root
of \eqref{lovepoly}, which will be negative since all $\alpha_k \ge 0$. As $\alpha_0$ increases (or as the pressure gets larger) 
the polynomial \eqref{lovepoly} retains its shape but increases everywhere by the same amount, and the smallest root moves increasingly closer to the location of the first local minimum of \eqref{lovepoly}. 
  The maximal pressure (maximal value of $\alpha_0$) occurs when this root is degenerate with the location of this minimum.
 
It is possible to obtain the thermodynamic quantities characteristic of the black hole solution as functions of the horizon radius $r_+$ 
(determined as the largest root of $f(r_+)=0$)
without explicitly solving for the metric function $f$ using the Hamiltonian formalism \cite{Cai:2003kt,Kastor:2011qp}. In Planckian units 
$l^2_P =\frac{G\hbar}{c^3} $ \cite{Kubiznak:2016qmn}, they are
\begin{align}
    M&=\frac{\Sigma^{(+1)}_{d-2}(d-2)}{16 \pi G_N}\sum_{k=0}^{N} \alpha_k r_+^{d-1-2k} + \frac{\Sigma^{(+1)}_{d-2} Q^2}{2(d-3)r_+^{d-3}},\\
    T&= \frac{1}{4\pi r_+ D}\left[ \sum_k \alpha_k \frac{d-2k-1}{r_+^{2(k-1)}} - \frac{8\pi G_N Q^2}{(d-2) r_+^{2(d-3)}}\right], \\
    S&= \frac{\Sigma^{(+1)}_{d-2} (d-2)}{4 G_N} \sum_{k=0}^{N} \frac{k \alpha_k r_+^{d-2k}}{d-2k},\\
    \Phi &= \frac{\Sigma^{(+1)}_{d-2} Q}{(d-3) r_+^{d-3}},
\end{align}
with 
\be
    D=\sum_{k=1}^N k \alpha_k r_+^{-2(k-1)}.
\ee

To avoid solutions with naked singularities in higher order Lovelock theories \cite{Myers:1988ze}, we will limit all $\alpha_k$ to be strictly nonnegative. We also set $\alpha_1 = G_N =1$ to recover general relativity when $\alpha_k =0$ for $k \ge 2$.

\section{multi-critical Phase Transitions}

The most straightforward way to study thermodynamic phase transitions is via the Gibbs free energy $G$
\be
    G=M-TS=G(T; P,Q,\alpha_1,\dots,\alpha_N)
\ee
 whose global minimum as a function of the temperature $T$ 
 determines the thermodynamically stable state of the system for a given choice of fixed $(P,Q,\alpha_1,\dots,\alpha_N)$.

Swallowtails in the Gibbs free energy are indicative of first-order phase transitions, with such transitions taking place at the intersection point of the swallowtail.  For $n$ distinct phases, $n-1$ swallowtails must be present. Previously, two swallowtails were shown to be possible in $d=6$ Gauss-Bonnet (second-order Lovelock) gravity, but more swallowtails were not seen in third-order Lovelock gravity \cite{Frassino:2014pha}. 
When the intersection points of these two swallowtails are coincident, we have the simplest example of a multicritical point, namely a triple or tricritical  point, again 
seen in $d=6$ Einstein-Gauss-Bonnet gravity \cite{Frassino:2014pha,Wei:2014hba}.

\begin{figure}
	\includegraphics[width=0.483\textwidth]{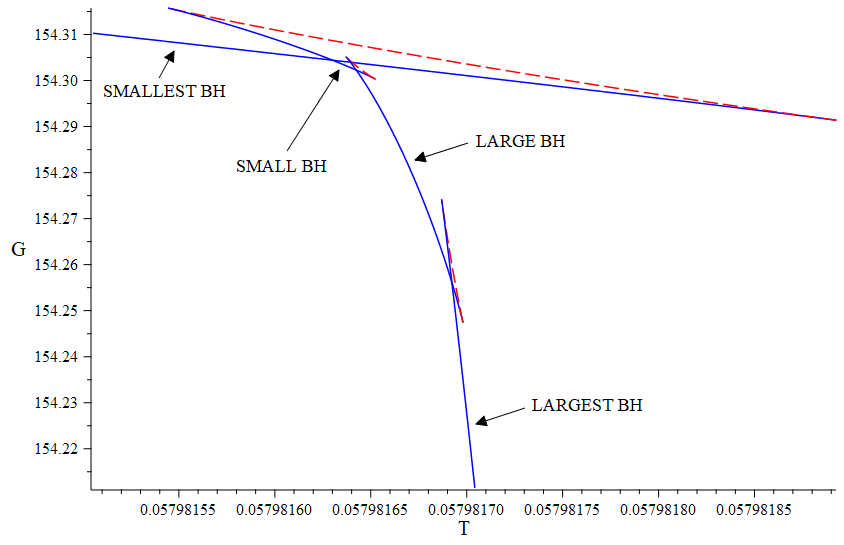}

	\caption{\textbf{$G$-$T$ plot: Three phase transitions at different temperatures of a charged Lovelock black hole.} Here $d=10$, $N=4$, $Q\approx 1.3467$, $\alpha_0 = 0.00368366$, $\alpha_2 \approx 26.367$, $\alpha_3 \approx 132.46$, $\alpha_4 \approx 106.91$. Three swallowtails in the Gibbs free energy for a charged black hole are seen to occur at different temperatures. Dashed curves indicate negative specific heat.}
	\label{fig:4ll-gt-separated} 
\end{figure}

\begin{figure}
	\includegraphics[width=0.483\textwidth]{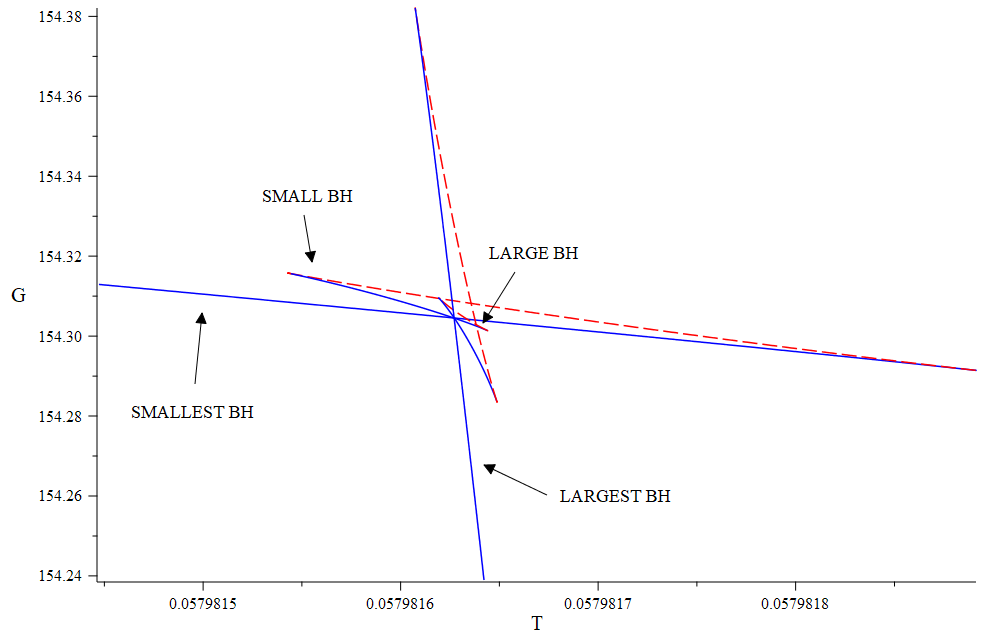}

	\caption{\textbf{$G$-$T$ plot: Quadruple point of a charged Lovelock black hole.} Here  $d=10$, $N=4$, $Q\approx 1.3467$, $\alpha_2 \approx 26.367$, $\alpha_3 \approx 132.46$, $\alpha_4 \approx 106.91$. At $\alpha_{0_q} \approx 0.0036835$ ($P_q \approx 0.0052762$), three swallowtails in the Gibbs free energy merge at a quadruple point. Dashed curves indicate negative specific heat.}
	\label{fig:4ll-gt-quad} 
\end{figure}

We therefore begin by seeking quadruple points, and find for a charged black hole system  in fourth-order Lovelock gravity in 10 spacetime dimensions 
  that indeed there can be  four distinct black hole phases.  There are  three swallowtails in the Gibbs free energy,
illustrated in
figure~\ref{fig:4ll-gt-separated}. For a different value of $\alpha_0$, the intersection points of the swallowtails merge at a quadruple point, as shown in figure~\ref{fig:4ll-gt-quad}.

\begin{figure}
	\includegraphics[width=0.483\textwidth]{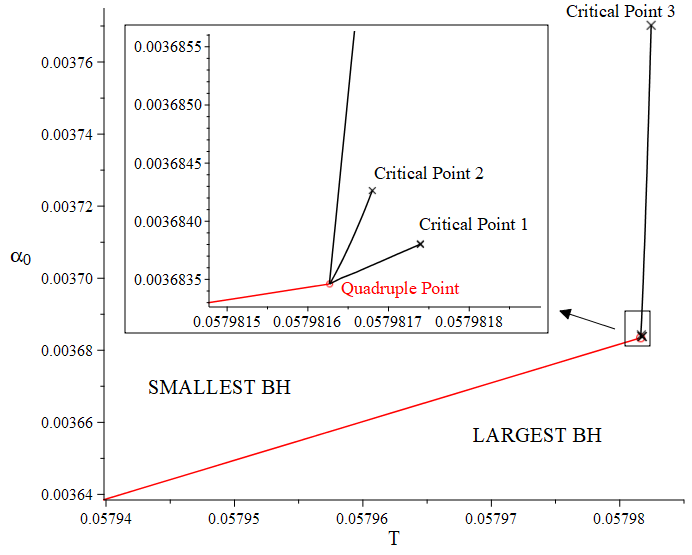}

	\caption{\textbf{$\alpha_0$-$T$ plot: Phase diagram of a charged Lovelock black hole.} Again, $d=10$, $N=4$, $Q\approx 1.3467$, $\alpha_2 \approx 26.367$, $\alpha_3 \approx 132.46$, $\alpha_4 \approx 106.91$. Four coexistence curves merge at $\alpha_{0_q} \approx 0.0036835$ ($P_q\approx 0.0052762$). For $\alpha_0 \in (\alpha_{0_q}, \alpha_{0_{c1}} \approx 0.0036838)$, two new intermediate phases emerge between the smallest and largest black holes, separated by first order phase transitions. These phase transitions terminate at critical points 1 and 2. The coexistence curve between the smallest and larger black holes (which are indistinguishable for $\alpha_0$ above the second critical point) terminates at $\alpha_{0_{c3}} = 0.0037702$. One stable phase transition exists between the largest and smallest black holes in the range $\alpha_0 \in (0,\alpha_{0_q})$. (red curve)}
	\label{fig:4ll-pt-quad} 
\end{figure}

As the pressure is decreased below the quadruple point pressure $P_q$, all smaller swallowtails become ``embedded'' in one large swallowtail, signifying the only stable phase transition for this pressure range. This large swallowtail is always present in the range $P\in (0, P_q)$, but the smaller unstable swallowtails can disappear at critical points. On the other hand, increasing the pressure above $P_q$ separates the three swallowtails, and three stable first order phase transitions are seen as the temperature of the black hole varies. The swallowtails diminish in size as we continue increasing the pressure, and ultimately disappear at their respective critical points. Figure~\ref{fig:4ll-pt-quad} shows the coexistence curves characteristic of the system.
\begin{figure}
	\includegraphics[width=0.483\textwidth]{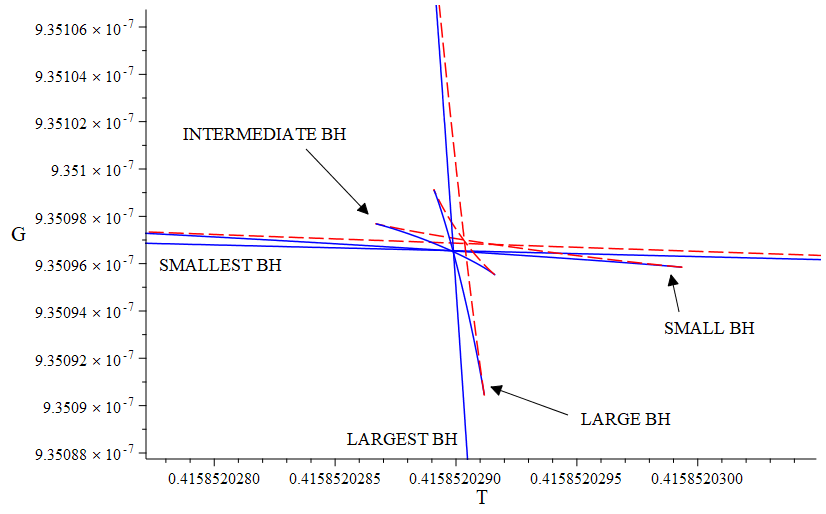}

	\caption{\textbf{$G$-$T$ plot: Quintuple point of a charged Lovelock black hole.} Here $d=14$, $N=6$, $Q\approx 1.87512377858\times 10^{-9}$, $\alpha_2 \approx 1.34267219033$, $\alpha_3 \approx 0.4720037602$, $\alpha_4 \approx 0.051852300491$, $\alpha5_5 \approx 0.0016877966277$, $\alpha_6 \approx 0.0000112370896883$. At $\alpha_{0_q} \approx 0.087519906865$, four swallowtails in the Gibbs free energy merge at a quintuple point. Dashed curves indicate negative specific heat.}
	
	\label{fig:6ll-gt-quin} 
\end{figure}
Higher order multi-critical points of this type can be constructed by adding more Lovelock constants. We explicitly show a quintuple point in $d=14$ sixth-order Lovelock gravity in figure~\ref{fig:6ll-gt-quin}. The behaviour here is very much the same as for the previously discovered multi-critical points \cite{Tavakoli:2022kmo,Wu:2022bdk}: multiple liquid/gas-like transitions with no solid/liquid analogue in the form of a Hawking-Page transition. The HP transition is forbidden in these systems due to respective conservation of charge and angular momentum. 
 
  Lovelock gravity can introduce new thermodynamic conjugate pairs without adding charge or angular momentum.  We find a novel type of multi-criticality involving an HP transition in neutral Lovelock black holes, allowing for arbitrarily many distinct phases.  This generalizes the  HP triple point first seen in exotic Lovelock black holes \cite{Hull:2021bry}.
  
\begin{figure}
	\includegraphics[width=0.48\textwidth]{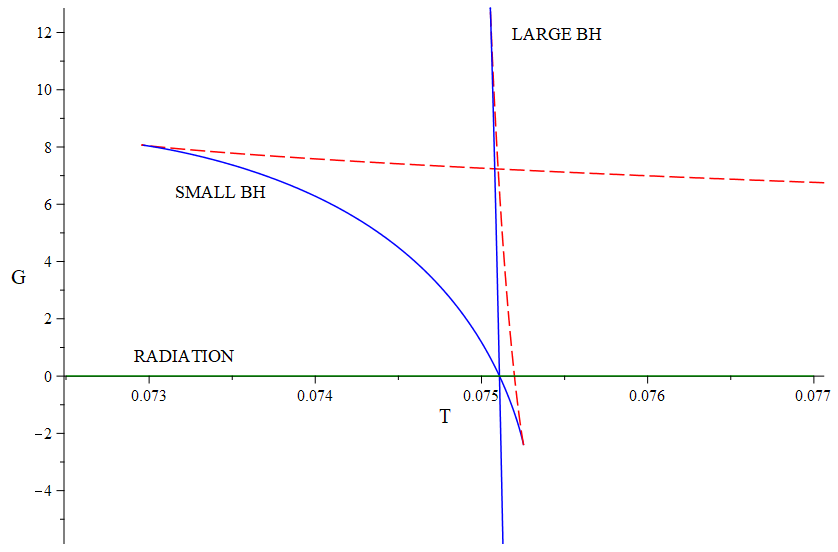}

	\caption{\textbf{$G$-$T$ plot: HP Triple point of a neutral black hole.} $d=8$, $N=3$, $\alpha_2 \approx 7.122112$, $\alpha_3 \approx 7.116266$. At $\alpha_0 \approx 0.00972336$, the swallowtail indicating the first order phase transition between large BH and small BH (blue curves) merge with the HP transition swallowtail. The green line $G=0$ represents thermal radiation. Dashed curves indicate negative specific heat.}
	\label{fig:3ll-gt-rad-tri} 
\end{figure}

\begin{figure}
	\includegraphics[width=0.48\textwidth]{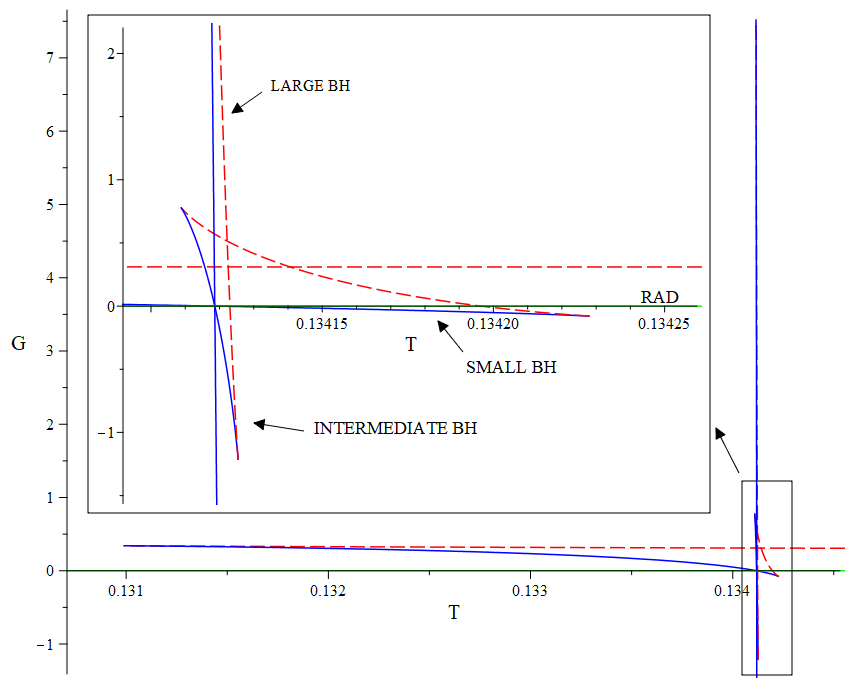}

	\caption{\textbf{$G$-$T$ plot: HP quadruple point of a neutral Lovelock black hole.} $d=12$, $N=5$, $\alpha_2 \approx 8.314253$, $\alpha_3 \approx 16.694181$, $\alpha_4 \approx 8.081339$, $\alpha_5 \approx 0.850375$. At $\alpha_0 \approx 0.0123316$, two swallowtails indicating first order phase transitions between large, intermediate, and small BHs (blue curves) merge with the HP transition at $G=0$. The green line $G=0$ represents thermal radiation. Dashed curves indicate negative specific heat.}
	\label{fig:5ll-gt-rad-quad} 
\end{figure}

  Figure~\ref{fig:3ll-gt-rad-tri} shows an HP triple point of radiation and small/large BHs in third-order Lovelock gravity in $d=8$,  whereas in figure~\ref{fig:5ll-gt-rad-quad} we depict an HP quadruple point of radiation and small/intermediate/large BHs in 5th-order Lovelock gravity in $d=12$.  In both of these cases, the number of Lovelock constants needed is more than that of their charged counterparts as the $(Q,\Phi)$ conjugate pair  no longer contributes to the extended phase space.

\begin{figure}
	\includegraphics[width=0.48\textwidth]{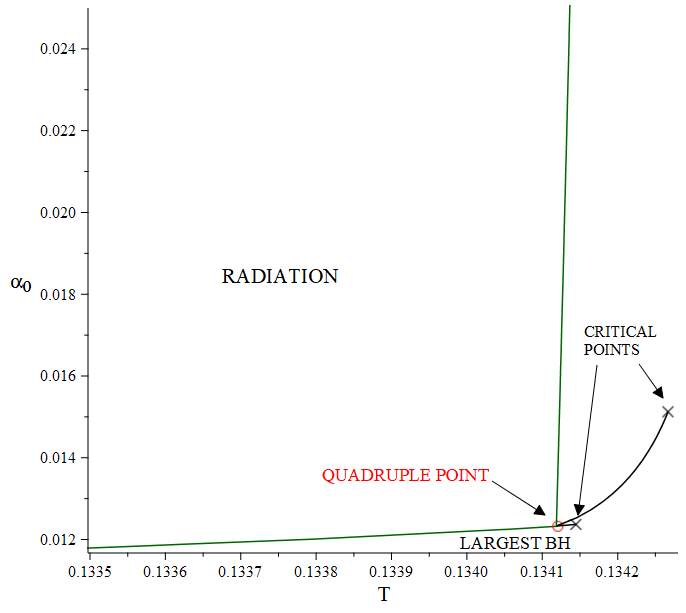}

	\caption{\textbf{$\alpha_0$-$T$ plot: Phase diagram of a neutral Lovelock black hole.} $d=12$, $N=5$, $\alpha_2 \approx 8.314253$, $\alpha_3 \approx 16.694181$, $\alpha_4 \approx 8.081339$, $\alpha_5 \approx 0.850375$. Four coexistence curves merge at $\alpha_{0_q} \approx 0.0123316$. For $\alpha_0 \in (\alpha_{0_q}, \alpha_{0_{c1}}=0.01237046)$, two new phases emerge between thermal radiation and the largest black hole phase, separated by first order phase transitions. These phase transitions terminate at their  respective critical points. The coexistence curve between radiation and black hole phases (which are indistinguishable for $\alpha_0$ above the second critical point) does not reach a critical point, but rather stops at the maximal pressure. One stable phase transition exists between the thermal radiation and the largest black hole in the range $\alpha_0 \in (0,\alpha_{0_q})$. The green curve indicates an  HP transition.}
	\label{fig:5ll-pt-rad-quad} 
\end{figure}

Uncharged black holes behave similarly to charged ones with one notable exception. As the pressure ($\alpha_0$) is raised, the free energy curve continues to have a zero at finite temperature; hence the HP transition does not disappear, and   its coexistence curve extends to the maximal pressure. This is analogous to the HP transition in Einstein gravity, which has an infinite coexistence curve
corresponding to a solid/liquid transition \cite{Kubiznak:2014zwa}.
 All other swallowtails terminate at their corresponding critical points for pressure values above the multi-critical pressure. On the other hand, reducing the pressure below the multi-critical pressure causes all phase transitions between BH phases to be ``embedded'' in the HP cusp formed by a green and a blue curve, driving the HP transition to be the only stable phase transition for $P\in(0,P_{mc})$. Figure~\ref{fig:5ll-pt-rad-quad} 
 summarizes this behaviour with a phase diagram for the system shown in figure~\ref{fig:5ll-gt-rad-quad}.

We observe multiple phase transitions only for a small set of Lovelock couplings, and finding appropriate values that can support multi-critical phenomenon proves to be challenging. We employ a method that focuses on temperature as a function of the horizon radius, for each extremum in $T(r_+)$ corresponds to a cusp in the Gibbs free energy plot \cite{Wu:2022bdk}. In general, $T$ is a rational function in $r_+$, and is continuous on $(0,\infty)$ for positive $\alpha_k$. For $n-1$ swallowtails ($n$ distinct phases), $2n-2$ extrema are required, and for the swallowtails to occur in the same temperature range,   a line of constant   temperature $T=T^*$ must intersect $T(r_+)$ $2n-1$ times:
\be
T\left (r_+^{(i)} \right) = T^* \text{ for } 1 \le i \le 2n-1.
\ee
This system of $(2n-1)$ equations is then solved for the $(N+2)$ unknowns ($T^*, Q, \alpha_0, \alpha_2, \dots,\alpha_N)$, indicating that $N \ge 2n-3$; two added coupling constants for each new phase.

However, not all $2n-1$ intersect conditions need to be enforced for charged black holes. With the $Q$ term in the denominator $T(r_+)$, we have the limits

\begin{equation}
    \lim_{r_+ \to 0^+} T(r_+) = -\infty, \quad \lim_{r_+ \to \infty} T(r_+) = \infty
\end{equation}
for positive $Q$. The continuity of $T$ on $r_+ \in (0,\infty)$ restricts the number of extrema to be even. Hence requiring that there are $(2n-2)$ $T^*$-intersects also guarantees the existence of a $(2n-1)$-th intersect. For $n-1$ swallowtails ($n$ distinct phases), we therefore only require $N\ge 2n-4$. 
Further micro-adjustments can be made to $r_+^{(i)}$ after the desired number of swallowtails is achieved in order to merge them at multi-critical points. 

In principle there is no limit to this procedure. 
Higher order multi-critical points can be obtained in higher curvature Lovelock gravity, with two new Lovelock constants guaranteeing a new phase. Note that the existence of $n$ phases is not sufficient to ensure the existence of an $n$-tuple point. With the assumption that all phases are separated by first order phase transitions, the $n-1$ coexistence curves can merge at multiple lower order multi-critical points rather than at a single $n$-tuple point. In fact, this is the case for most choices of $Q,\alpha_1,\dots,\alpha_N$.

The generalized Gibbs phase rule \cite{Sun:2021gpr} 
\be
\textsf{F}=\textsf{W}-\textsf{P}+1,
\ee
 relates the degrees of freedom $\textsf{F}$ of the system to the number of thermodynamic conjugate pairs $\textsf{W}$ and coexistent phases $\textsf{P}=n$. Both charged and neutral Lovelock black holes that solve \eqref{lovepoly}
  have $n-1$ degrees of freedom, one less than $\textsf{F}=n$ for black holes in non-linear electrodynamics \cite{Tavakoli:2022kmo}, 
  but much more than
 multiply rotating Kerr-AdS black holes in Einstein gravity, which have  $\textsf{F}=2$ and  require the least amount of thermodynamic conjugate pairs to achieve multi-criticality.
 In contrast to non-linear electrodynamics, where the temperature function is linear in the additional coupling constants, in Lovelock gravity 
 the temperature is a rational function of the couplings.  It may be that
 fewer Lovelock coupling constants are required to achieve comparable results. The fact that the jump from $n=2$ phases 
 in Einstein-Maxwell theory to 
$n=3$ requires only one additional Lovelock coupling (the GB coupling constant) provides some evidence for this. 
 Furthermore, black hole systems with $n=0$ degrees of freedom still remain to be observed, despite being fairly common in ordinary matter. 
 
\section{Conclusions}

Black Hole multicriticality is a generic feature of $N$-th order Lovelock gravity theories, with an $n$-tuple point possible provided $N \geq 2n-4$.  
By regarding the Lovelock coupling constants as thermodynamic parameters
\cite{Kastor:2010gq},  choices of the parameters exist that allow for $n$-tuple critical points.  These choices, while fine-tuned, do not constitute a set of measure zero in parameter space. While
tricriticality ($n=3$) has been known for quite some time \cite{Frassino:2014pha,Wei:2014hba}, we  have shown that both charged and neutral 
$N\geq 4$ Lovelock black holes exhibit multicriticality, with $n > 3$.

Both charged and neutral Lovelock black holes that solve \eqref{lovepoly}  have $n-1$ degrees of freedom, but charged black holes exhibit one greater degree of multicriticality than neutral black holes do at a given order $N$.   We have also found a new phenomenon of HP-multicriticality, in which several distinct phases of a neutral black hole are indistinguishable from thermal AdS at a particular value of pressure and temperature.

Multicomponent systems in nature, such as polymers and colloids, exhibit multicritical behaviour \cite{Sun:2021gpr,PhysRevLett.125.127803}. 
Our results indicate that black holes can 
behave in a likewise manner, 
providing further evidence that gravitational 
thermodynamics resembles chemical thermodynamics in all its aspects.  The implications of this for a quantum theory of gravity remain to be understood.

\section*{Acknowledgements}
\label{sc:acknowledgements}

This work is supported in part by the Natural Sciences and Engineering Research Council of Canada (NSERC).  Perimeter Institute and the University of Waterloo are situated on the Haldimand Tract, land that was promised to the Haudenosaunee of the Six Nations of the Grand River, and is within the territory of the Neutral, Anishnawbe, and Haudenosaunee peoples.

\bibliographystyle{JHEP}

\providecommand{\href}[2]{#2}\begingroup\raggedright\endgroup

\end{document}